\documentstyle[12pt,epsfig]{article}
\newcommand{\be}{\begin{eqnarray}}
\newcommand{\ee}{\end{eqnarray}}

\newcommand{\bi}{\begin{itemize}}
\newcommand{\ei}{\end{itemize}}

\begin{document}
\title{Non-local kinetic simulation of heavy ion collisions}
\author{Klaus Morawetz\\
Max-Planck-Institute for the Physics of Complex Systems,\\
Noethnitzer Str. 38, 01187 Dresden, Germany}
\date{}
\maketitle
\begin{abstract}
The numerical solutions of nonlocal and local Boltzmann kinetic
equations
for the simulation of peripheral and central heavy ion reactions are compared. 
The experimental finding of enhancement of mid-rapidity
matter shows the necessity to include nonlocal corrections. 
While the in--medium cross
section
changes the number of collisions and not the transferred energy,
 the nonlocal scenario changes the energy
transferred during collisions. The renormalisation of quasiparticle
energies by using the
Pauli--rejected collisions
leads
to a further enhancement of mid--rapidity matter distribution and is
accompanied by a
dynamical softening of the equation of state. 

 The simulation results are parameterised in terms
of time dependent thermodynamical variables in the Fermi liquid
sense. This allows one to discuss
dynamical trajectories in phase space. A combination of
thermodynamical observables is constructed which locates
instabilities and points of possible phase transition under 
iso-nothing conditions. Two different mechanisms of instability, a short time
surface--dominated instability and later a spinodal--dominated volume
instability is found. The latter one occurs only if the incident energies do
not exceed significantly the Fermi energy and might be attributed to spinodal
decomposition. 
\end{abstract}

\section{Introduction}

The formation of a neck--like structure in peripheral heavy ion
reactions and the impact on the fragmentation mechanism and production of
light charged particles has been discussed for a couple of years 
\cite{Stuttge92}-
\cite{SDCD97}. 
Theoretical investigations
suggest that the neck is not formed in usual heavy ion  simulations starting from the Landau equation \cite{PR87,DDM90,DRS91}
or BUU equations \cite{KH94,ACC95} including additional mean field fluctuations derived
in \cite{BV85,Fl89} and tested \cite{TV85}. The inclusion of
fluctuations in the Boltzmann (BUU) equation has been investigated
resulting in Boltzmann-Langevin pictures
\cite{ColDiT95,CTGMZW98,FCD98}-
\cite{BCR91}.

We will take here the point of view that the fluctuations should arise
by themselves in a proper kinetic description where all relevant
correlations 
are
included in the collision integral. The collision will then cause both
a dephasing and fluctuation by itself. This procedure without
additional assumption about fluctuations has been given by the
nonlocal kinetic theory \cite{SLM96,MLSCN98,LSM99} and applied
to heavy ion  collisions in \cite{MLSK98}-
\cite{MLNCCT01}. We claim
that the derived nonlocal off-set in the collision procedure induces
fluctuations in the density and consequently in the mean-field which
are similar to the one assumed in Boltzmann Langevin approaches above. 

Recent INDRA observation shows an enhancement of emitted matter in the
region of almost zero relative velocity which means that matter is
stopped
during the reaction and stays almost at rest \cite{B00,P00,G98,L00}.
This enhancement of
mid--rapidity distribution can possibly be associated with a pronounced neck
formation of matter.

\begin{figure}
\parbox[]{17.5cm}{
\parbox[]{8cm}{ 
\psfig{figure=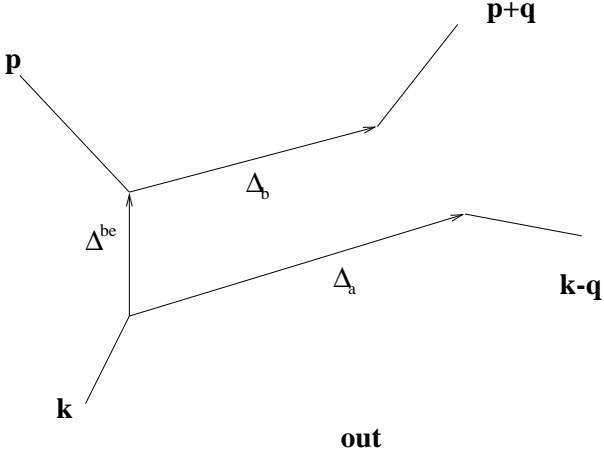,width=8cm}}
\hspace{1cm}
\parbox[]{7.5cm}{\vspace*{4cm} 
\caption{A nonlocal binary collision according to
  Eq. (\protect\ref{9}). Here $\Delta^{\rm be}=\Delta_2$,
  $\Delta_a=\Delta_3$ and $\Delta_b=\Delta_4-\Delta_2$ is used.
\label{soft}}
}}
\end{figure}

We use for the description the
nonlocal kinetic
equation \cite{SLM96,LSM97} for the one - particle distribution function
\begin{eqnarray}
\!\!\!&&{\partial f_1\over\partial 
t}+{\partial\varepsilon_1\over\partial {\bf k}}
{\partial f_1\over\partial 
{\bf r}}-{\partial\varepsilon_1\over\partial {\bf r}}
{\partial f_1\over\partial {\bf k}}
=
\sum_b\int{d{\bf p}d{\bf q}\over(2\pi)^5\hbar^7}\delta\left(\varepsilon_1
+
\varepsilon_2-\varepsilon_3-\varepsilon_4+2\Delta_E\right)
\left|{\cal T}_{ab}\right|^2
=\nonumber\\
&&\Bigl[f_3f_4\bigl(1-f_1\bigr)\bigl(1-f_2\bigr)-
\bigl(1-f_3\bigr)\bigl(1-f_4\bigr)f_1f_2\Bigr],
\label{9}
\end{eqnarray}
with Enskog-type shifts of arguments \cite{SLM96,LSM97}:
$f_1\equiv f_a({\bf k,r},t)$, $f_2\equiv 
f_b({\bf p,r}\!-\!\Delta_2,t)$,
$f_3\equiv 
f_a({\bf k\!-\!q}\!-\!\Delta_K,{\bf r}\!-\!\Delta_3,t\!-\!\Delta_t)$
, and
$f_4\equiv
f_b({\bf p\!+\!q}\!-\!\Delta_K,{\bf r}\!-\!\Delta_4,t\!-\!\Delta_t)$
. The
effective scattering measure, the ${\cal T}$-matrix is 
centred in all
shifts. The quasiparticle energy $\varepsilon$ contains 
the mean field
as well as the correlated self energy.
In agreement with \cite{NTL91,H90}, all gradient 
corrections are given
by derivatives of the scattering phase 
shift \mbox{$\phi={\rm Im\ ln}{\cal 
T}^R_{ab}(\Omega,{\bf k,p,q,}t,{\bf r})$},
\begin{equation}
\begin{array}{lclrclrcl}
\Delta_2&=&
{\displaystyle\left({\partial\phi\over\partial {\bf p}}-
{\partial\phi\over\partial {\bf q}}-{\partial\phi\over\partial {\bf k}}
\right)_{\varepsilon_3+\varepsilon_4}}&\ \
\Delta_3&=&
{\displaystyle\left.-{\partial\phi\over\partial {\bf k}}
\right|_{\varepsilon_3+\varepsilon_4}}&\ \
\Delta_4&=&
{\displaystyle-\left({\partial\phi\over\partial {\bf k}}+
{\partial\phi\over\partial {\bf q}}\right)_{\varepsilon_3+\varepsilon_4}}
\\ &&&&&&&&\\
\Delta_t&=&
{\displaystyle \left.{\partial\phi\over\partial\Omega}
\right|_{\varepsilon_3+\varepsilon_4}}&\ \
\Delta_E&=&
{\displaystyle \left.-{1\over 2}{\partial\phi\over\partial t}
\right|_{\varepsilon_3+\varepsilon_4}}&\ \
\Delta_K&=&
{\displaystyle \left.{1\over 2}{\partial\phi\over\partial {\bf r}}
\right|_{\varepsilon_3+\varepsilon_4}}.
\end{array}
\label{8}
\end{equation}
After derivatives, 
$\Delta$'s are evaluated at the energy shell 
$\Omega\to\varepsilon_3+
\varepsilon_4$. Neglecting these shifts the usual BUU 
scenario appears.
The scattering integral of the non-local kinetic equation derived in
\cite{SLM96,LSM97} corresponds to the picture of a collision as seen
in Figure~\ref{soft}.

Despite its complicated form it is possible to solve 
this kinetic equation 
with standard Boltzmann numerical codes and to implement 
the shifts \cite{MLSCN98,MT00,MLNCCT01}.
 To this end we have calculated the shifts for different 
realistic nuclear potentials
 \cite{MLSK98}. These shifts and the modifications of standard BUU or
 QMD code are available from the author. The numerical solution of the nonlocal 
kinetic equation 
has shown an observable effect in the dynamical particle 
spectra \cite{MLSCN98} as well as in the charge density distribution 
\cite{MLNCCT01}. 
The high energetic tails of the spectrum are enhanced due 
to more energetic two-particle collisions 
in the early phase of nuclear collision. Therefore the 
nonlocal corrections lead 
to an enhanced production of preequilibrium high 
energetic particles.

Besides the nonlocal shifts and cross section which has been
calculated from realistic potentials, the interaction affects, 
however, the free motion
of particles between individual collisions. The dominant effect is due 
to mean-field forces which bind the nucleus together, accelerating
particles close to the surface towards the centre. These mean-field forces are 
conveniently included via potentials of Skyrme type
\be
\varepsilon={p^2\over 2 m} + A \left ({n\over n_0}\right )+B \left
  ({n\over n_0}\right )^\sigma.
\label{hf}
\ee
For a derivation of collision integrals and the Skyrme potential
(\ref{hf}) from the same microscopic footing, see \cite{M00}.

Beside forces, the interaction also modifies the velocity with which a
particle of a given momentum propagates in the system. This effect is 
known as the mass renormalisation. A numerical implementation of the
renormalised mass is rather involved since a plain use of the
renormalised mass instead of the free one leads to incorrect currents.
Within the Landau concept of quasiparticles, this problem is cured by 
the back flow, but it is not obvious how to implement the back flow 
within the BUU simulation scheme. In our studies, we circumvent the 
problem of back flows using explicit zero-angle collisions to which we 
add a non-local correction. One can show that indeed the non-local
shifts create just the dynamical mass renormalisation if used for the
Pauli-rejected events. The incorporation of a nonlocal jump without
performing finite angle collisions for such events correspond exactly to the 
quasiparticle and dynamical mass renormalisation \cite{MLNCCT01}.

Finally, we would like to comment on properties of the proposed
simulation scheme. The renormalisation depends on the distribution of
particles in surrounding medium. It has four nice properties: (i) the
renormalisation vanishes as the local density goes to zero, (ii) the
renormalisation vanishes when a high temperature closes the Luttinger
gap because all collisions will be at finite angles, (iii) the
anisotropy of the quasiparticle velocity in a presence of a non-zero
current in medium is automatically covered, and (iv) the backflows
connected to the mass renormalisation are covered because both
particles jump keeping the centre of mass fixed. Last but not least,
the simulation does not require to introduce new time-demanding
procedures, one can simply use the scattering events which are merely
rejected in standard simulation codes by Pauli-blocking.

\section{Numerical results}

Let us discuss the proposed correction to the local and ideal (no
quasiparticle renormalisation) Boltzmann (BUU) simulation. First we
introduce the pure nonlocal corrections and then we discuss the
quasiparticle renormalisation.

The evolution of the density can be seen in the corresponding left
pictures of figure~\ref{midrap08} for the
BUU (left panel) and nonlocal scenario (middle panel) as well as the additional quasiparticle
renormalisation (right panel). We see that the nonlocal scenario leads
to a longer and more pronounced neck formation between $200-240$fm/c
while the BUU breaks apart already at $200$fm/c.

\begin{figure}[]
\centerline{\psfig{file=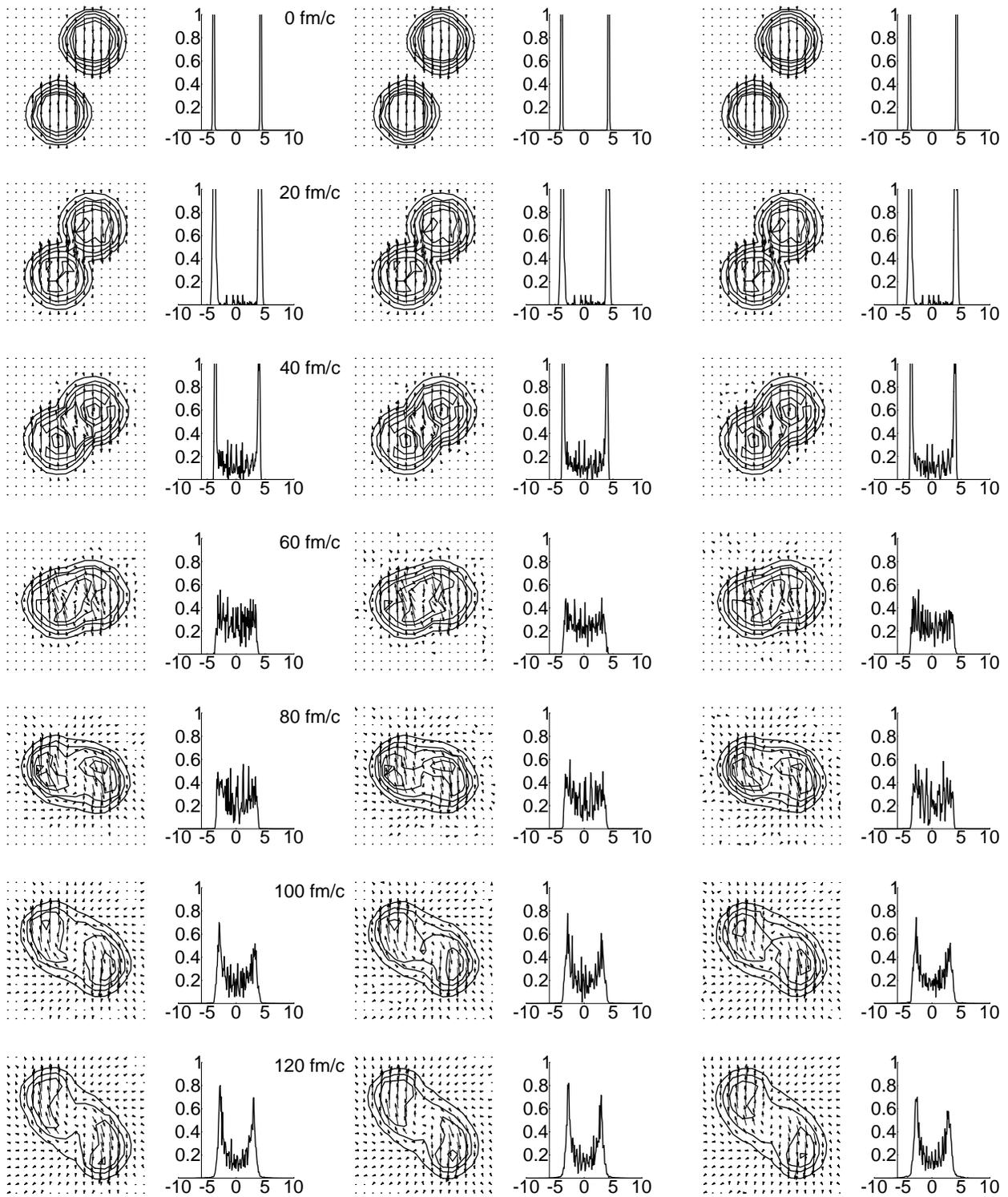,height=20cm,angle=0}}
\caption{First part.}
\end{figure}
\setcounter{figure}{1}
\begin{figure}[]
\centerline{\psfig{file=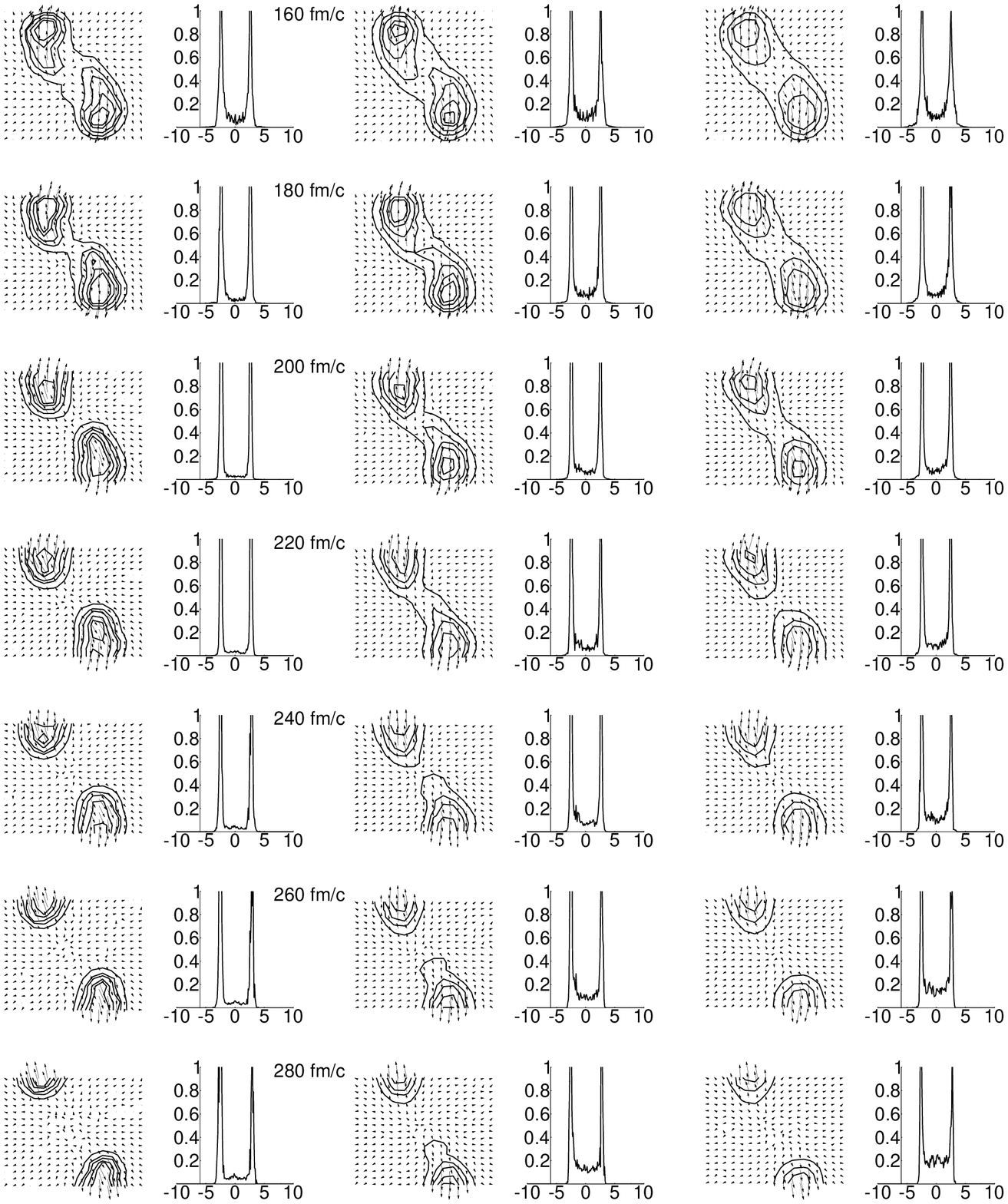,height=20cm,angle=0}}
\caption{The time evolution of $^{181}_{73}Ta + ^{197}_{79}Au$ 
collisions at $E_{lab}/A = 33$~MeV and $8$~fm impact parameter in  
the local BUU (left), non-local BUU (middle), and the non-local BUU 
with quasiparticle renormalisation (right). Plots in the first column 
show the $(x-z)$-density cut of $30$~fm$\times 30$~fm where $Ta$ as
projectile 
comes from below.
The mass momenta are shown by arrows. The corresponding second columns
give the charge density distribution versus relative velocity in
$cm/ns$ where the target like distribution of $Au$ is on the left and
the projectile like distributions of $Ta$ on the right.\label{midrap08}
}
\end{figure}

The question arises whether this pronounced neck formation and
increase of mid-rapidity distribution is simply occurring
by more collisions and corresponding correlations as called in--medium
effect. This is however only the case for smaller impact
parameters \cite{G98}.  We see
in the next figure \ref{n} the number of collisions per time for the
 simulation where
in the local BUU scenario the cross section has been doubled. 
The
number of
collisions is visibly enhanced by doubling the cross section while
for the nonlocal scenario we get only a slight enhancement at the
beginning and later even lower values with respect to local BUU. The
latter fact comes from the earlier decomposition of matter in the
nonlocal scenario. 
\begin{figure}[h]
\parbox[]{17.5cm}{
\parbox[]{8cm}{ 
\psfig{file=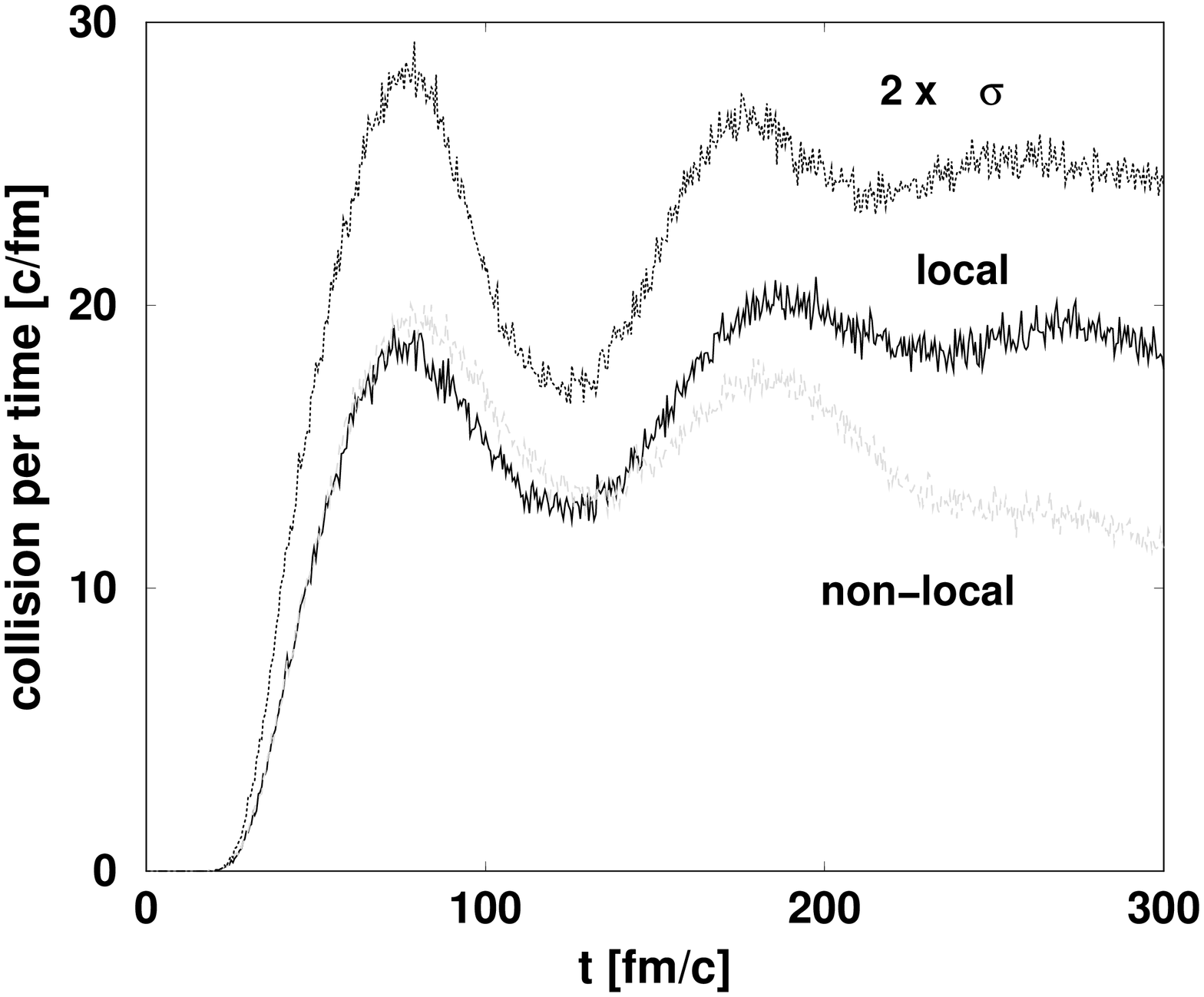,height=8cm,angle=-90}
}
\hspace{1cm}
\parbox[]{7.5cm}{\vspace*{2cm} 
\caption{The time evolution of the number of nucleon collisions for
$Ta+Au$ at $E_{lab}/A = 33$ MeV and different impact parameter in  
the BUU (thick black line), non-local kinetic model (broken line) and 
the local BUU with a doubled 
cross section (thin dark line). The impact parameter is 8~fm.\label{n}}
}
}
\end{figure}

\begin{figure}[h]
\parbox[]{17cm}{
\parbox[]{8cm}{ 
\psfig{file=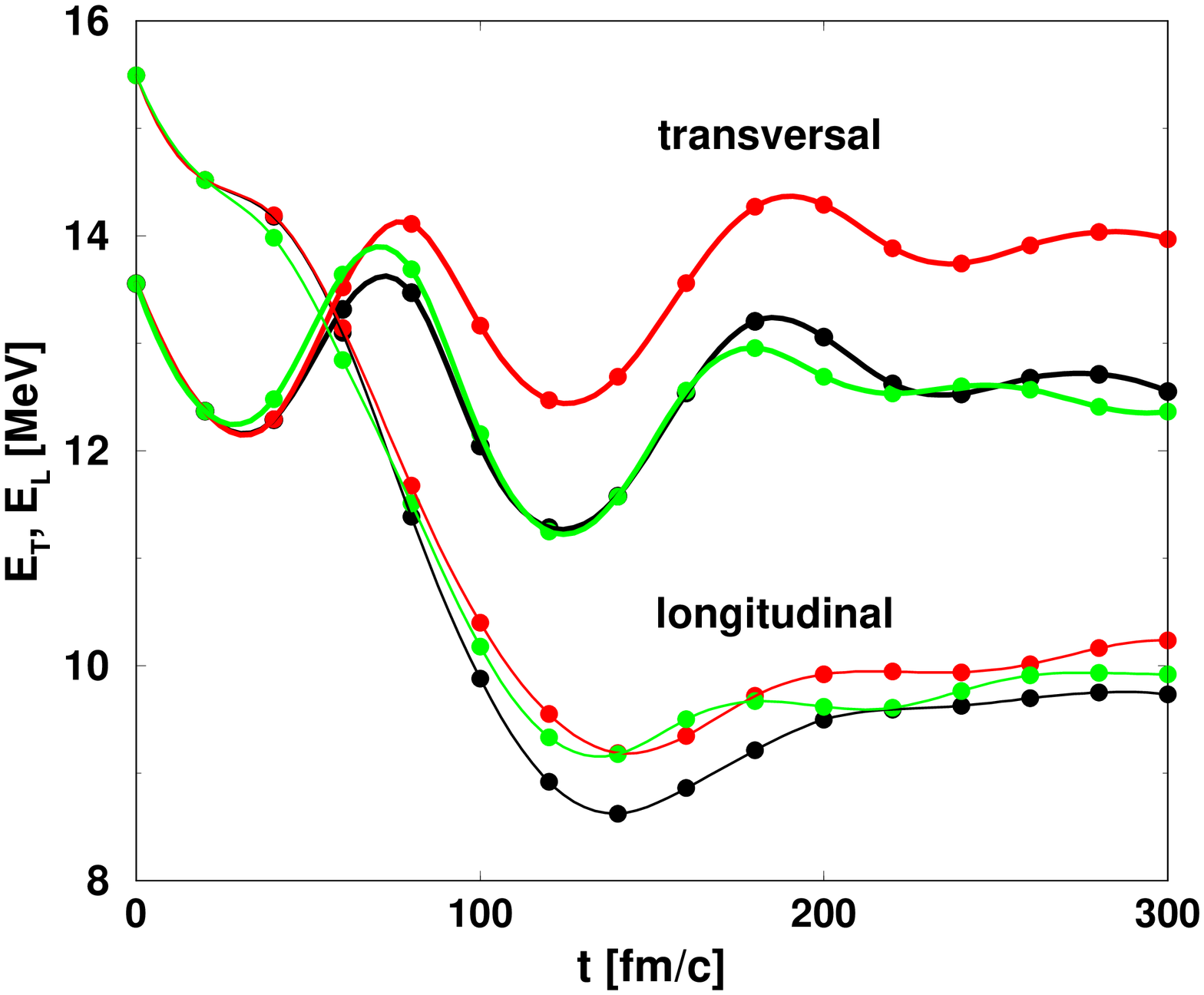,height=8cm,angle=-90}
}
\hspace{1cm}
\parbox[]{7.5cm}{\vspace*{1cm} 
\caption{The time evolution of the longitudinal (thin lines) and
  transverse energy (thick lines)
  including Fermi motion  of nucleon collisions for
  $Ta + Au$ at
$E_{lab}/A = 33$ MeV in  the BUU (black
  line),
nonlocal kinetic equation (light line) and  the local BUU
but with twice the cross section (medium line).
\label{e}
}
}}
\end{figure}

The
corresponding transverse and kinetic energies in figure \ref{e}
($8$fm impact parameter) show that the transverse and longitudinal energy is almost not changed
compared with local BUU. Oppositely the nonlocal scenario leads to an
increase of transverse energy of about $2$MeV and about $1$MeV in
longitudinal energy. We conclude that the increase of cross section
leads to a higher number of collisions but not to more dissipated
energy while the nonlocal scenario does not change the number of
collisions much but the energy dissipated during the collisions. 
Returning to the discussion of pronounced neck formation in figure
\ref{midrap08} above we see now that the quality
 rather than the quantity of collisions is what produces the neck. 

\begin{figure}[h]
\parbox[]{17cm}{
\parbox[]{10cm}{ 
\psfig{file=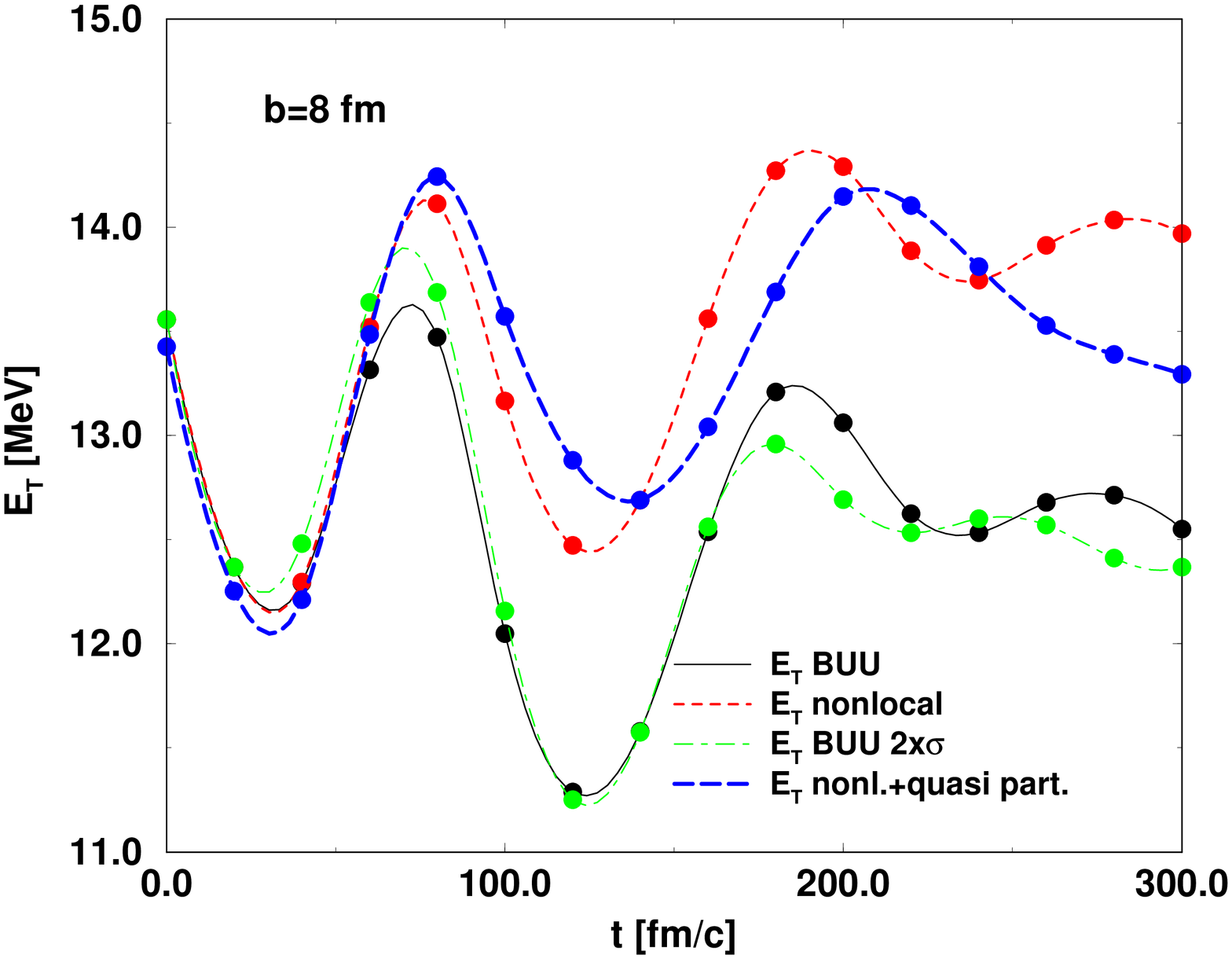,height=10cm,angle=-90}
}
\hspace{1cm}
\parbox[]{5.5cm}{\vspace*{1cm} 
\caption{The time evolution of the transverse energy
  including Fermi motion for
  $Ta + Au$ at
$E_{lab}/A = 33$ MeV and  $8$fm impact parameter in  the BUU (black
  line),
nonlocal kinetic equation (dashed line), the local BUU
but twice cross section (dashed dotted line) and the nonlocal scenario with
  quasiparticle renormalisation (long dashed line).
\label{e08}
}
}}
\end{figure}

Now we can proceed and discuss the charge matter distribution with
respect to the velocity.
We plotted in figure \ref{midrap08}
also the normalised charge distribution versus velocity and see that
after $160$fm/c we have an appreciable higher mid--rapidity
distribution for the nonlocal scenario (mid panel) than the BUU
(left panel). Together with the observation that for nonlocal scenario
we have a pronounced neck formation we see indeed that the neck
formation is accompanied with high mid-velocity distribution of
matter.
We see in figure \ref{midrap08}
(right panel) that the mid--rapidity distribution of matter is once more
enhanced for quasiparticle renormalisation in comparison to the
nonlocal 
scenario without quasiparticle
renormalisation. The detailed comparison of the
time evolutions of the transverse energy for $8$fm impact parameter
can
be seen in figure \ref{e08}.
We recognise that the transverse energies including quasiparticle
renormalisation are
similar to the nonlocal scenario and higher than the BUU or BUU with
twice the cross section. However please remark that the
period of oscillation in the transverse energy which corresponds to a
giant resonance becomes larger for the case with quasiparticle
renormalisation. Since therefore the energy of this resonance
decreases we can conclude that the (nuclear) compressibility has been decreased
by the quasiparticle renormalisation. Sometimes this quasiparticle
renormalisation has been introduced by momentum dependent
mean-fields.
The effect is known to soften the
equation of state. We see here that we get a dynamical quasiparticle
renormalisation and a softening of equation of state.
This softening of equation of state is already slightly remarkable
when
the nonlocal
scenario is compared with BUU. With additional quasiparticle
renormalisation we see that this is much pronounced.

We want to repeat that the dynamical quasiparticle
renormalisation which leads to a softening of the equation of state  
enhances the mid rapidity distribution. In
contrast a mere soft static parametrisation of the mean-field does not
change the mid-rapidity emission appreciably \cite{G98}.

\subsection{Comparison with experiments}

The BUU simulations will now be compared to one experiment performed
with INDRA at GANIL, the $Ta + Au$ collision at $E_{lab}/A = 33$ MeV.
\begin{figure}
\parbox[]{17cm}{
\parbox[]{10cm}{ 
\psfig{file=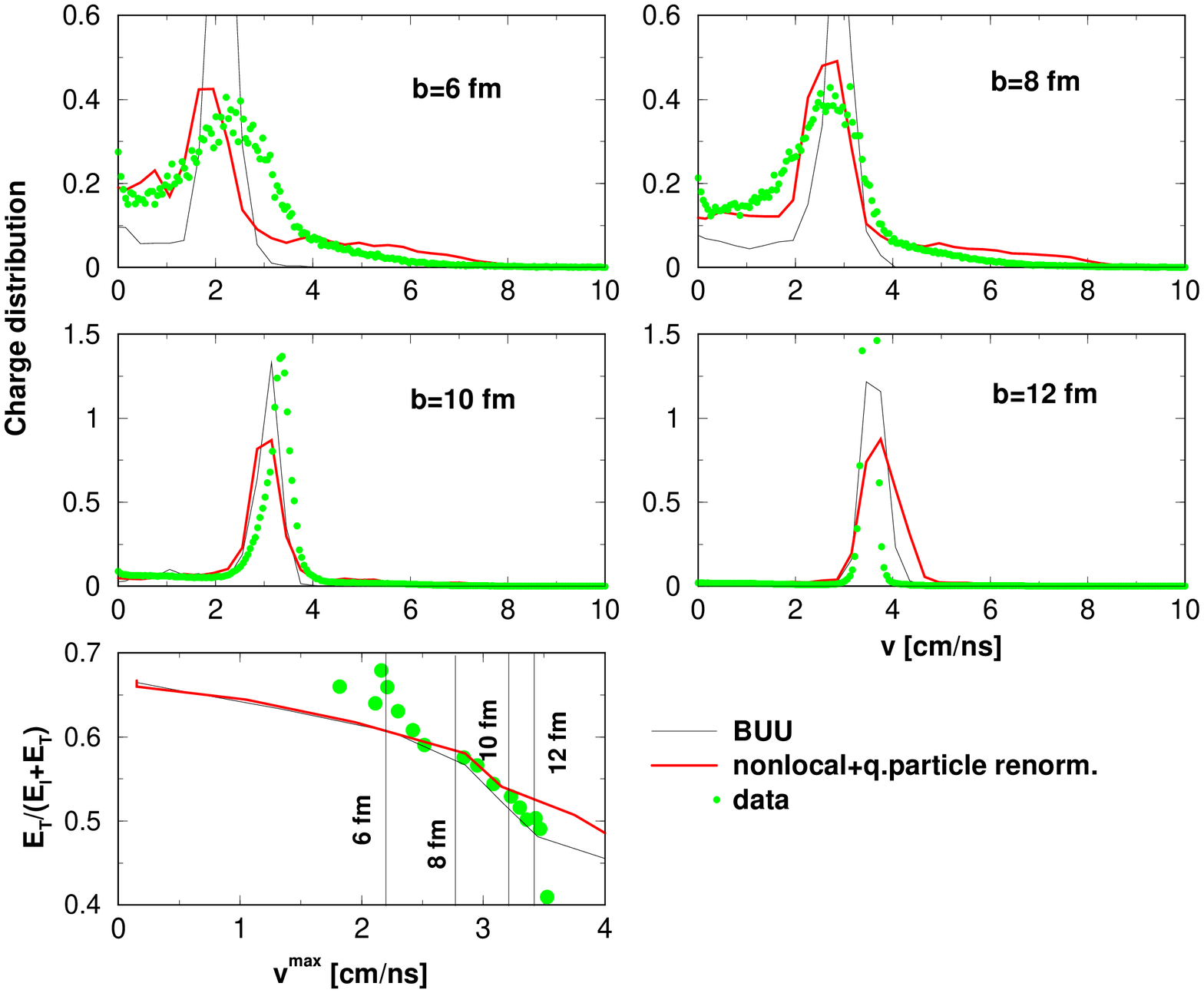,height=10cm,angle=-90}
}
\hspace{1cm}
\parbox[]{5.5cm}{
\caption{The experimental charge distribution of matter (dotted line)
versus velocity in  comparison within  the BUU (thin solid line) and
the nonlocal model with quasiparticle
renormalisation (thick line). The maximum velocity versus ratio of
  longitudinal to total kinetic energy  of $Ta + Au$ collisions at
$E_{lab}/A = 33$ MeV is given below. The selected experimental cuts are given by dots.\label{ch_p}
}
}}
\end{figure}
For the identification with experimental data we select events which
show a clear one fragment structure. This correspond to events where
we have clear target and projectile like residues. Since the used
kinetic theory is not capable to describe dynamical fragment formation
we believe that these events are the ones which are at least
describable within our frame. Next we use impact parameter cuts with
respect to the transverse energy since this shows in all simulations a
fairly good correlation to the impact parameter. In our numerical
results we see almost linear correlations between impact parameter,
maximal velocity and the convenient ratio between transverse and
total kinetic energy as seen in figure \ref{ch_p}.

For each selected experimental transverse energy bin we can plot now
the maximum velocity versus the ratio of the transverse to kinetic
energy. We see in figure \ref{ch_p} that the numerical velocity
damping agrees with the experimental selection
only for very peripheral collisions. For such events we plot in
the figure \ref{ch_p} the charge density distribution and compare the experiment with
the simulation. These charge density distributions have been obtained using the
procedure described in reference \cite{L00}. The Data are represented by light grey
points, the standard BUU calculation by the thin line and the non-local BUU
with quasiparticle renormalisation calculation by the thick line.  A
reasonable agreement is found for the nonlocal
scenario including quasiparticle renormalisation while simple BUU
fails to reproduce mid--rapidity matter.

\section{Nonequilibrium Thermodynamics}

In order to understand the different thermodynamics induced by
nonlocal corrections and therefore virial correlations,
we multiply the kinetic equation with $1,{\bf p},\varepsilon$ and obtain the
balance for the particle density $n$, the momentum density $J$ and the
energy density ${\cal E}$, see details in \cite{MT00}. 
Please note that besides the mean field (\ref{hf}) we have also a Born
correlation term, see \cite{MK97,M00},

The correlational parts of the density, pressure and energy are
arising from genuine two-particle correlations beyond Born
approximation which are also derived from the balance equations of
nonlocal kinetic equations \cite{SLM96,LSM97}. It has been shown that they
establish the complete conservation laws. 
While these correlated parts are present in the numerical results and
can be shown to contribute to the conservation laws we will only discuss the
thermodynamical properties in terms of quasiparticle quantities to
compare as close as possible with the mean field or local BUU
expressions. The discussion of these correlated two - particle
quantities is devoted to a separate consideration.

From the distribution function $f({\bf p,r},t)$ the local density, current
and energy densities
are given by
\be
\left (\matrix{
{n}({\bf r},t)
\cr
{\bf J}({\bf r},t)
\cr
{\cal E}_K({\bf r},t)
}\right )&=&
\int{d {\bf p}\over (2\pi \hbar)^3}
\left (\matrix{
1
\cr
{\bf p}
\cr
{p^2/2 m}
} \right )
{f({\bf p,r},t)}
\label{var}
\ee
which are computed directly from the numerical solution of the kinetic
equation in terms of test particles. Please note that the above
kinetic energy includes the Fermi motion.

The global variables per particle number like kinetic energy, Fermi
energy and collective energy are obtained by spatial integration
\be
\left (\matrix{
{\cal E}_K(t)
\cr
{\cal E}_F(t)
\cr
{\cal E}_{\rm coll}(t)
}\right )
&=&
\int d{\bf r} 
\left (\matrix{
{\cal E}_K({\bf r},t)
\cr
{\cal E}_f[{n}({\bf r},t)]{n}({\bf r},t) 
\cr
{{J}({\bf r},t)^2 /m \,{n}({\bf r},t)} 
}\right ){/\int d{\bf r} \, {n}({\bf r},t)}
\ee
where we have used the local density approximation \cite{SHJGRSS89}.
Now we adopt the picture of Fermi liquid theory which connects the
temperature with the kinetic energy as
\be
{\cal E}_K(t)=\frac 3 5 {\cal E}_F(t)+ {\cal E}_{\rm coll}(t) +
{\pi^2\over 4 {\cal E}_F(t)} {
  T(t)}^2
\label{t}
\ee
from which we deduce the global temperature. The definition of
temperature is by no means obvious since it is in principle an
equilibrium quantity. One has several possibilities
to define a time dependent equivalent
temperature which
should approach the equilibrium value when the system approaches
equilibrium. In \cite{GW00,GWF00} the
definition of slope temperatures has been discussed and compared to local space
dependent temperature fits of the distribution function of
matter. This seems to be a good measure for higher energetic
collisions in the relativistic regime. 
Since we restrict us here to collisions in the Fermi energy domain and do not want to add coalescence models we will not use
the slope temperature. Moreover we define the global temperature in
terms of global energies which are obtained by local quantities rather
than
by defining a local temperature itself. This has the advantage that we do
not consider local energy fluctuations but only a mean evolution of temperature.

The mean field part of the energy is given by
\be
U(t)&=&{\cal E}^{\rm qp}(t)-{\cal E}_K(t)-{\cal E}^{\rm
  Born}(t)
={\displaystyle{\int d{\bf r} \left (A{{ n({\bf r},t)}^2 \over 2 n_0}
    +B{ {n({\bf r},t)}^{s+1} \over (s+1) n_0^s}\right )}/
\displaystyle{\int d{\bf r} \, {n}({\bf r},t)}}
\ee
from which one deduces the pressure per particle
\be
P(t)&=&\frac 2 3 ({\cal E}_K(t)-{\cal E}_{\rm coll}(t))+ \frac 4 3 {\cal E}^{\rm
  Born}(t)
+{\displaystyle{\int d{\bf r} \left (A{ { n({\bf r},t)} \over 2 n_0}
    +B{ s \,{ n({\bf r},t)}^{s} \over (s+1) n_0^s}\right )} / \displaystyle{\int d{\bf r}\, { n}({\bf r},t) }}.
\ee

In order to compare now the local BUU with the nonlocal BUU scenario we
consider the energy which would be the total energy in the local BUU
without Coulomb energy
\be
{\cal E}(t)={\cal E}_K(t)-{\cal E}_{\rm coll}(t)+U(t).
\label{ee}
\ee

This expression does not contain the two - particle correlation energy
which is zero for BUU and the Coulomb energy. The reason for
considering this energy for dynamical trajectories is that we want to
follow the trajectories in the picture of mean field and usual spinodal
plots.

To define the density exhibits to some extend a problem. 
Depending on the considered volume sphere
we obtain different global densities. We follow here the point of view
that the mean square radius will be used as a sphere 
to define the global density. This is also supported by the
observation that the mean square radius follows the visible
compression. 

\subsection{Iso-nothing conditions in equilibrium}

\begin{figure}[h]
\centerline{\parbox{17cm}{
\parbox{8cm}{
\psfig{file=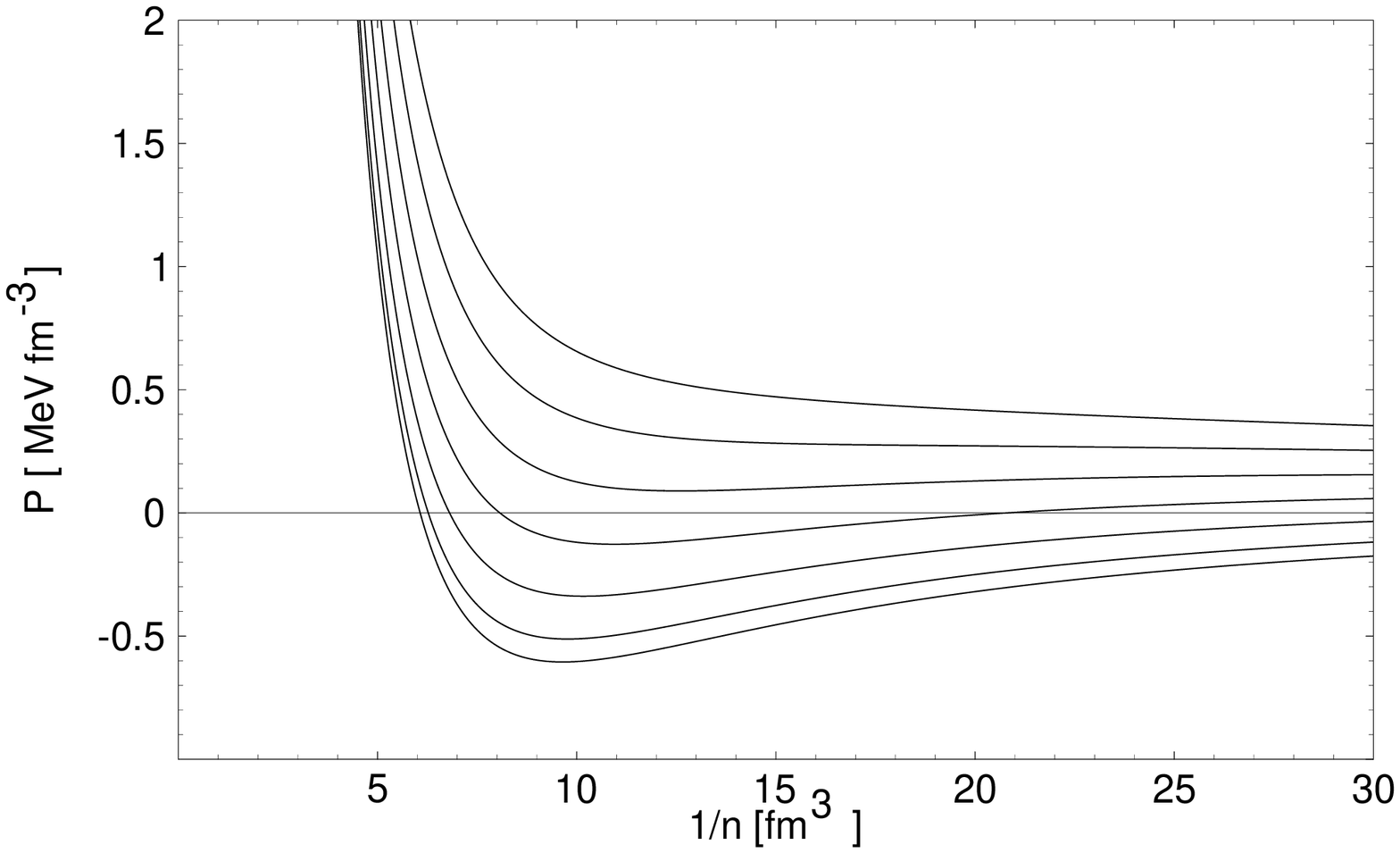,width=8cm}
}
\hspace{0.5cm}
\parbox{8cm}{
\psfig{file=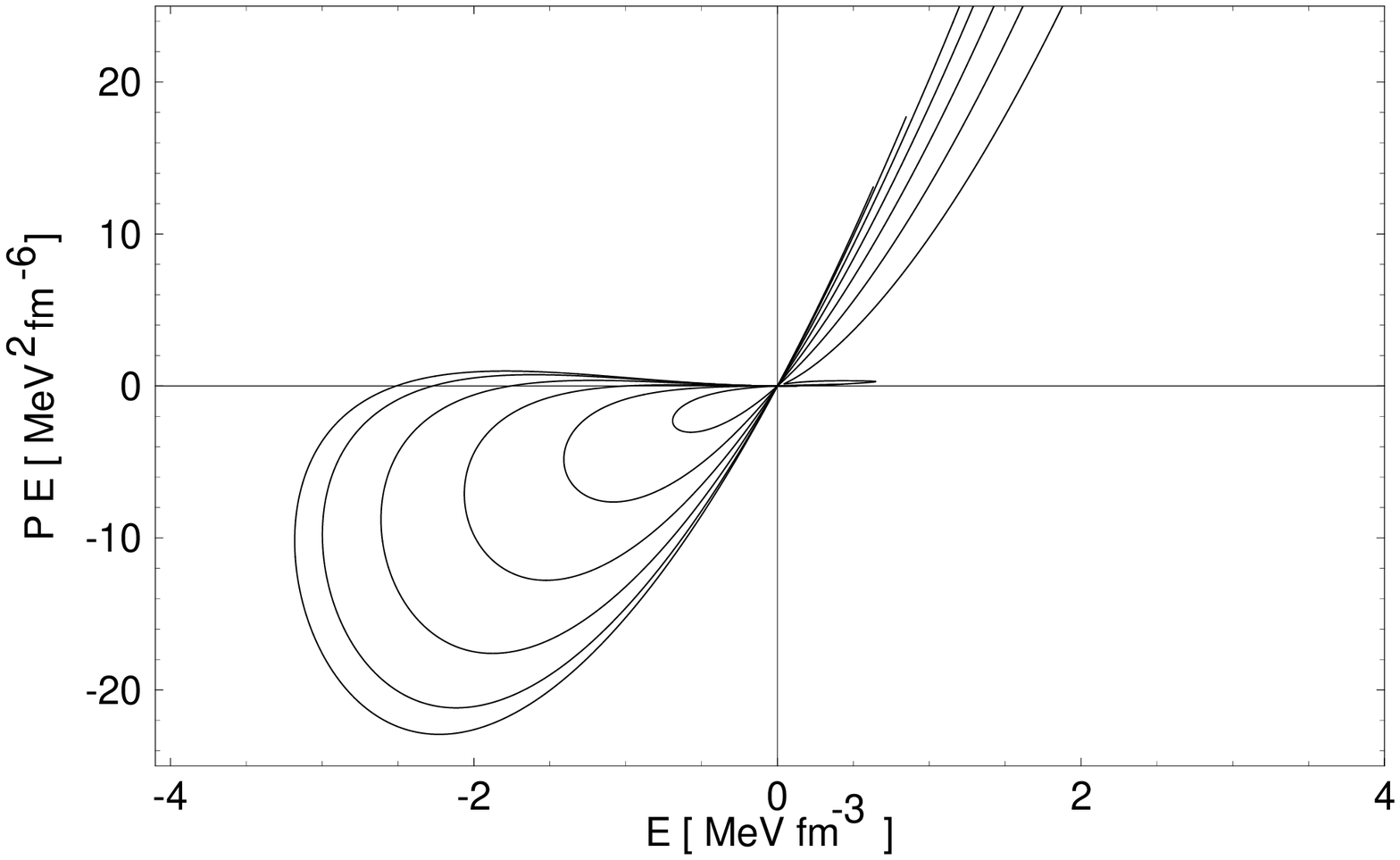,width=8cm}}
}}
\parbox[]{17cm}{
\parbox[]{8cm}{
\psfig{file=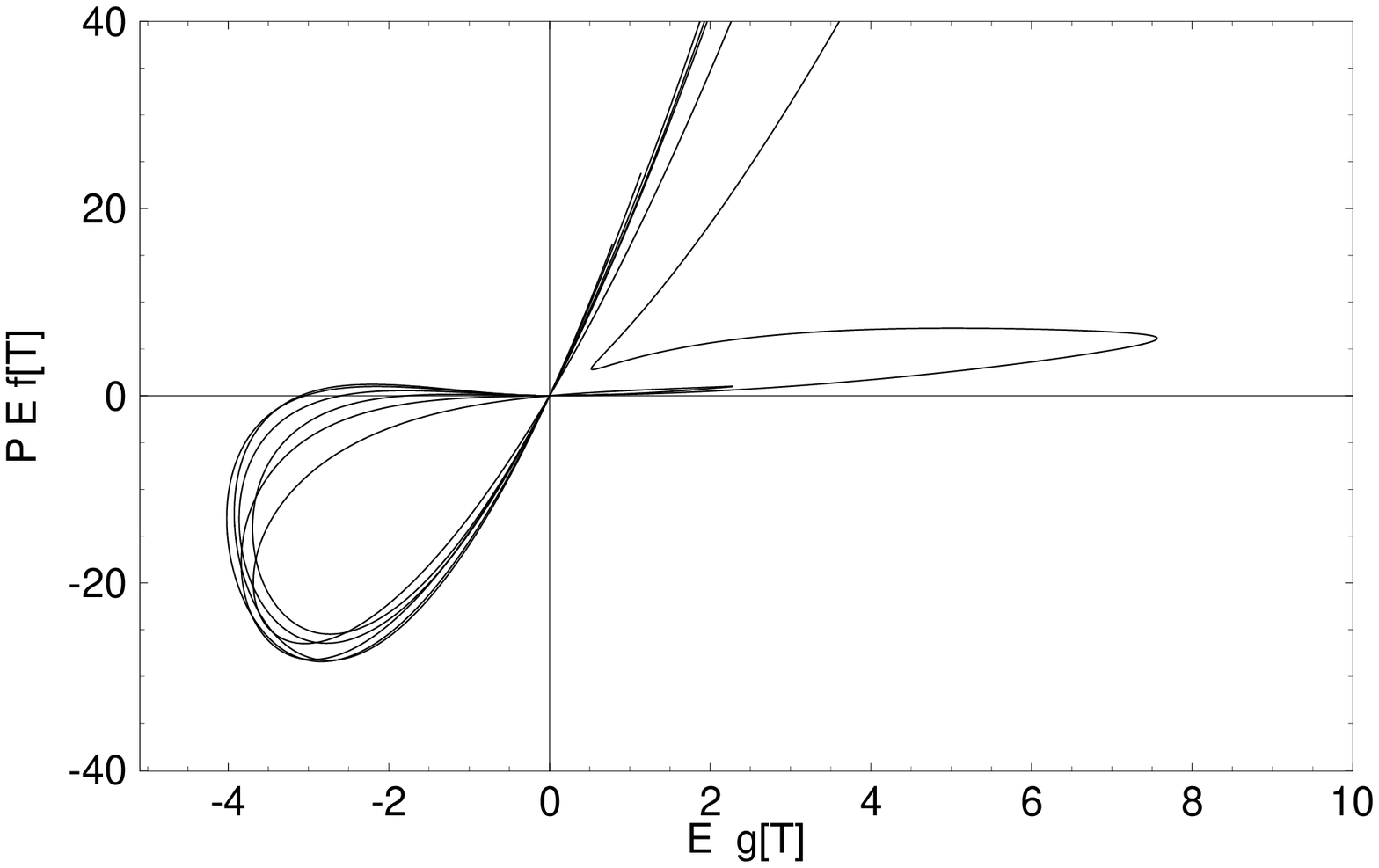,width=8cm}
}
\hspace{1cm}
\parbox[]{7.5cm}{
\caption{The isotherms for the pressure density versus volume and for
  the product of pressure and energy density versus energy density
  (above). The temperatures are $T=1,4,7,10,13,16,19$MeV.
The isotherms of the product of pressure and energy density
  versus energy density scaled by a temperature polynomial $f(T)$,
  $g(T)$ (below). All critical isotherms
collapse on one line in the left lower quarter.\label{1}} 
}}
\end{figure}

Let us first recall the figures of mean field isotherms in
equilibrium. We obtain the typical van der Waals curves in figure
\ref{1}. Since we have neither isothermal nor isochoric nor isobaric
conditions in simulations, shortly since we have iso-nothing conditions, we
have to find a representation of the phase transition curves which are
independent of temperature but which reflect the main features of phase
transitions. This can be achieved by the product of energy and
pressure density versus energy density in figure \ref{1} below. This
plot shows that all instable isotherms exhibit a minimum in the left
lower quarter. There the energy is negative denoting bound state
conditions but
the pressure is already positive which means the system is
unstable. The first isotherm above the critical one does not touch
this quarter but remains in the right upper quarter where the energy
and pressure are both positive and the system is expanding and
decomposing unboundly. The left upper quarter denotes negative pressure
and energy indicating that the system is bound and stable.

In order to achieve now a temperature independent plot we scale both
axes of figure \ref{1} (below) with a temperature dependent polynomial
and achieve that all critical isotherms are collapsing on one curve in
the left lower quarter. 
The first isotherms above
the critical one does not enter the left lower quarter. We consider
this scaling as adequate for iso-nothing conditions. A phase
transition should be possible to observe if there occurs a minimum in
the left half of this plot at negative energies. 

The idea of plotting combinations of pressure and energy is
similar to the one of softest point \cite{HS95} in analysing QCD phase
transitions. There the simple pressure over energy ratio leads to a
temperature independent plot due to ultrarelativistic energy -
temperature relations. In our case we have a Fermi liquid behaviour at
low temperatures and  the scaling temperature dependent polynomials,
$f$ and $g$,
can be found in \cite{MT00} 
which are producing a temperature independent plot in figure \ref{1}.

\subsection{Simulation results}

\begin{figure}
\parbox[]{17cm}{
\parbox[]{10cm}{
\psfig{file=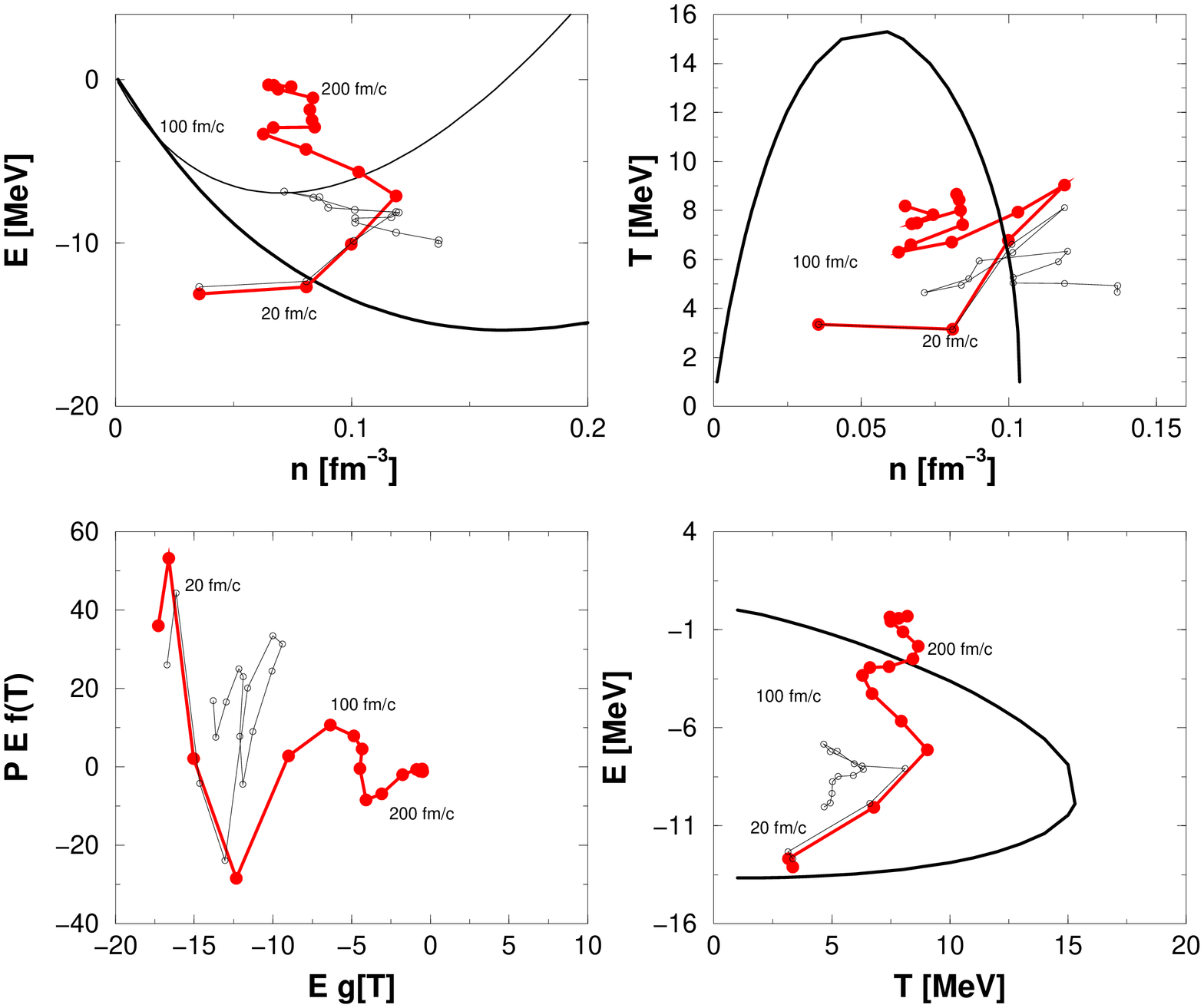,height=10cm,angle=-90}
}
\hspace{0.2cm}
\parbox[]{6.5cm}{
\caption{The dynamical trajectories of the energy (\protect\ref{ee}),
  density and
  temperature (i\protect\ref{t}) in the nonlocal (gray thick) and in
  the local
  BUU (black thin) scenario. The considered reaction is $^{129}Xe$ on $^{119}Sn$ at $25$MeV
  lab energy. The dots mark the times in steps of $20$fm/c up to
  total of $300$fm/c. To guide the eye the zero temperature mean field
  energy (thick line) and the pressure (thin line) is plotted in the upper
  left picture and in the right figures the
  spinodal line for infinite matter is given. The scaled
  combinational plot analogous to figure \protect\ref{1} is given in
  the left lower plane.\label{xesn25}}
}}
\end{figure}

Let us now inspect the dynamical trajectories for the above defined
temperature, density and energy. In figure \ref{xesn25} we have
plotted the dynamical trajectories for a charge - symmetric reaction
of $Xe$ on $Sn$ at $25$MeV lab - energy. The solution of the nonlocal
kinetic equation is
compared to the local BUU one. One sees in the temperature versus density
plane that the point of highest compression is reached around
$60$fm/c with a temperature of $9$ MeV. 

After this point of highest
overlap or fusion phase we have an
expansion phase where the density and temperature is decreasing. While
the compression phase is developing similar for the BUU and for the nonlocal
kinetic equation we see now differences in the development. First the
temperature of the nonlocal kinetic equation is around $2$MeV higher than
the local BUU result. This is due to the release of correlation
energy into kinetic energy which is not present in the local BUU
scenario. After this expansion stage until times of $120$fm/c we see that the
BUU trajectories come to a rest inside the spinodal region while the
nonlocal scenario leads to a further decay. This can be seen by
the continuous decrease of density and increase of
energy. Since matter is more decomposed with the nonlocal kinetic equation
we also heat the system more due to Coulomb acceleration. This
leads to the enhancement of temperature compared to BUU. An
oscillating behaviour occurs at later
times which reflects an interplay between short range correlation and long
range Coulomb repulsion. The decomposition leads almost to free
gaseous matter after $300$fm/c as can be seen in the energy versus
density plot. 

Please note that although the trajectories seem to
equilibrate inside the spinodal region when one considers the temperature
versus density plane, we see that in the corresponding energy versus
temperature plane the trajectories travel already outside the
spinodal region. This underlines the importance to investigate the region of
spinodal decomposition in terms of a three dimensional plot instead
of a two dimensional one like in the recently discussed caloric curve
plots. Different experimental situations lead to different curves as
long as the third coordinate (pressure or density) remains
undetermined.

The iso-nothing plot analogous to the left lower plot of figure \ref{1} 
shows that the point
of highest compression is linked to a first instability seen as a
pronounced minimum of the trajectory in the left quarter. If we do not
exceed the Fermi energy domain, this is
connected with a pronounced surface emission and connected with
anomalous velocity profiles
\cite{MTP00}. We will call this phase {\it surface emission}
instability further on.
At $180$ fm/c we see a second minimum which is taking place inside
the spinodal region. This instability we might now attribute to spinodal
decomposition since the trajectories developing slower and remain
inside the spinodal region. The BUU shows the same qualitative minima
but the
matter rebounds and the trajectories move towards negative energies
again. In opposition the nonlocal scenario leads to a further
decomposition of matter as described above.

\begin{figure}
\parbox[]{17cm}{
\parbox[]{10cm}{
\psfig{file=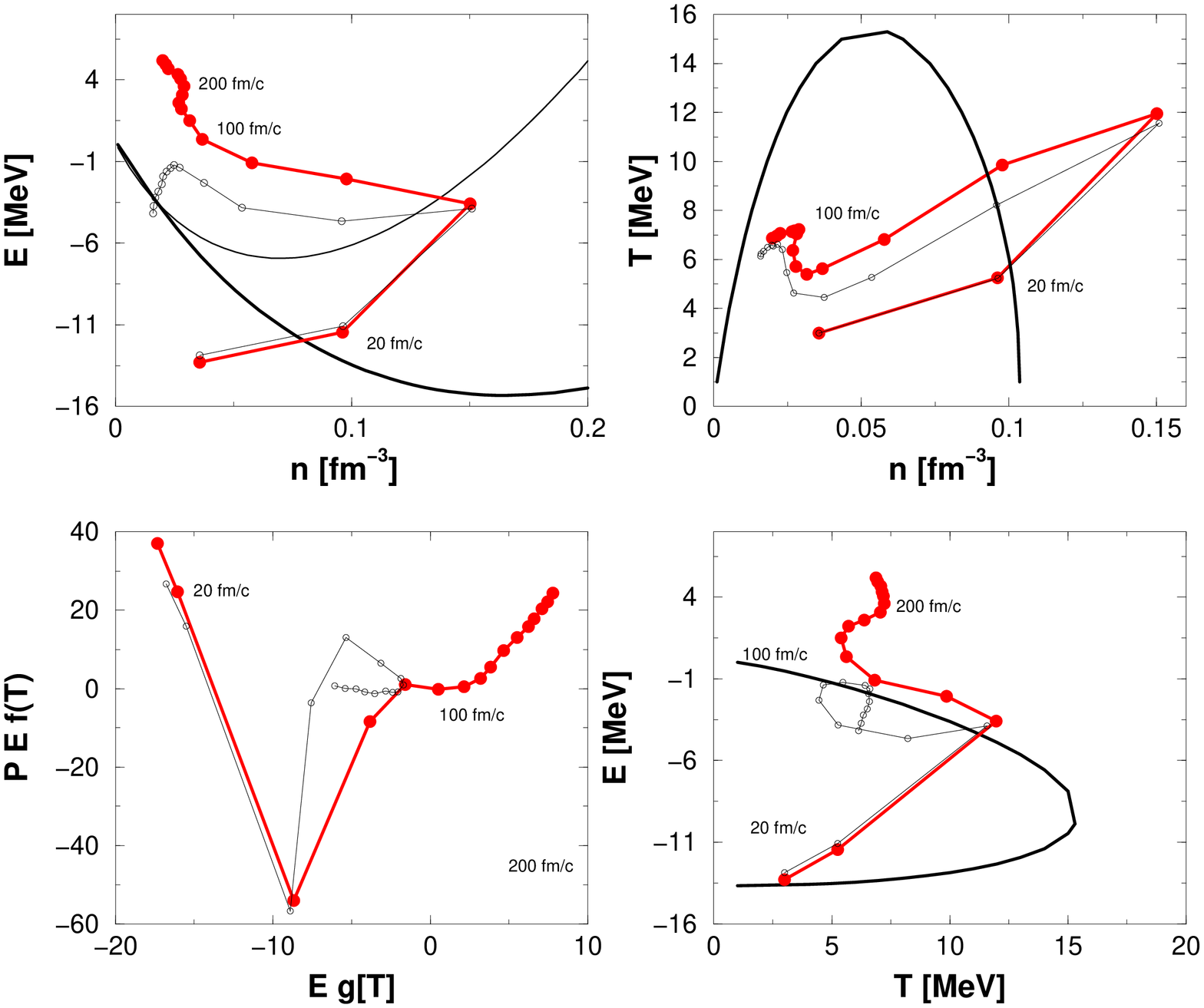,height=10cm,angle=-90}
}
\hspace{0.2cm}
\parbox[]{6.5cm}{
\vspace{7cm}
\caption{The same figure as \protect\ref{xesn25} but for $50$MeV lab
  energy.\label{xesn50}}
}}
\end{figure}

\begin{figure}
\parbox[]{17cm}{
\parbox[]{10cm}{
\psfig{file=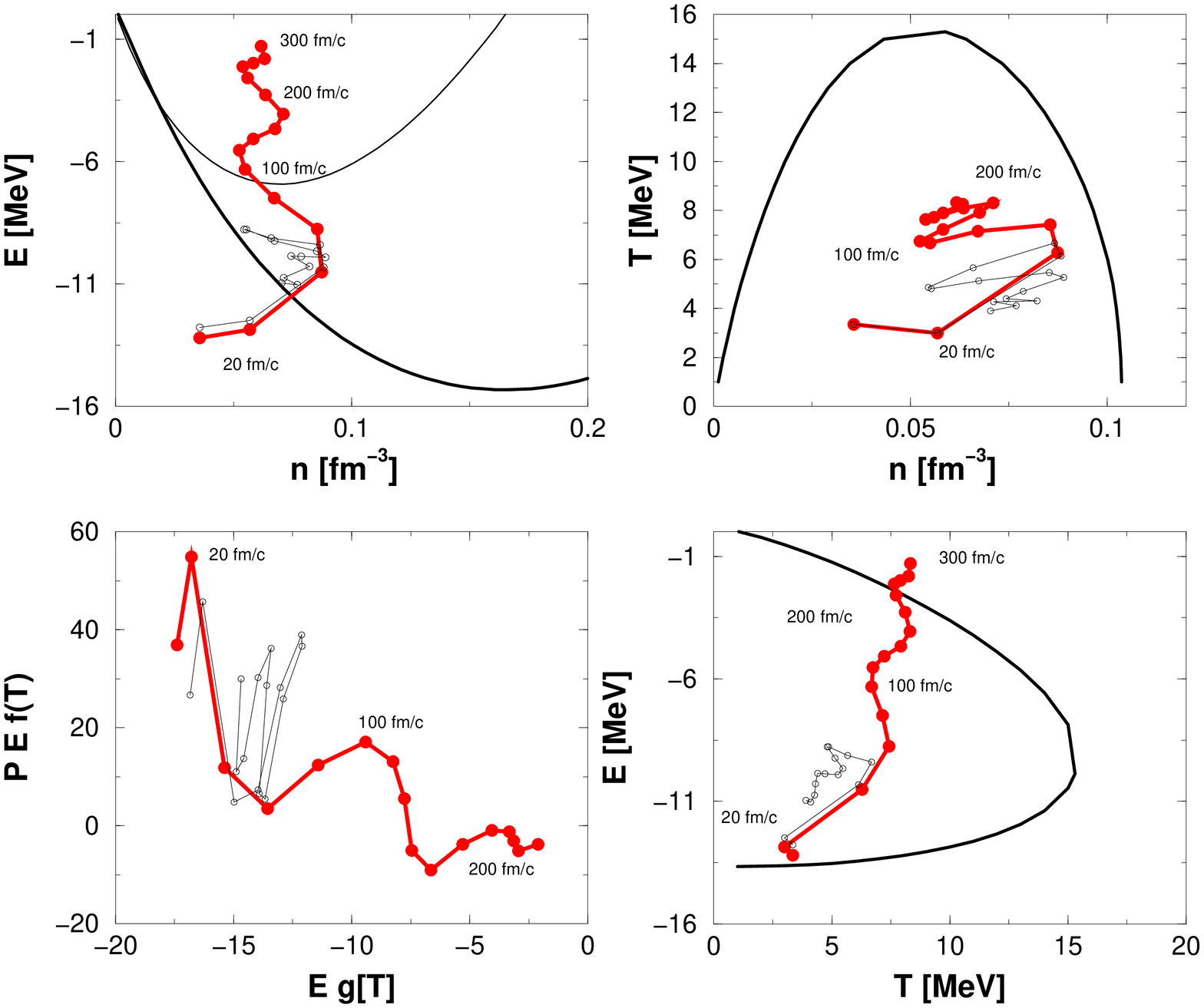,height=10cm,angle=-90}
}
\hspace{0.2cm}
\parbox[]{6.5cm}{
\vspace{6cm}
\caption{The same dynamical trajectories as in figure
  (\protect\ref{xesn25}) but for a reaction $^{56}Ni$ on $^{179}Au$ at $25$MeV
  lab energy. \label{niau25}}
}}
\end{figure}

In the next figure \ref{xesn33} we have plotted the same reaction as
in figure \ref{xesn25} but at higher energy of $33$MeV. We recognise a
higher compression density and temperature than compared to the lower
bombarding energy. Consequently the trajectories develop further towards
the unbound region of positive energy after $300$fm/c. While the first
surface emission instability is strongly pronounced we see that the
second minimum in the iso-nothing plot is already weaker indicating
that the role of spinodal decomposition is diminished. The
trajectories in the temperature versus density plot come still in the
spinodal region at rest but travel already outside the spinodal
region if the
energy versus temperature plot is considered. This shows that the
trajectories start to develop too fast to suffer
much spinodal decomposition. Oppositely, at energies around the Fermi
energy spinodal decomposition might be possible and is presumably
connected with anomalous velocity and density profiles \cite{MTP00}.
\begin{figure}
\parbox[]{17cm}{
\parbox[]{10cm}{
\psfig{file=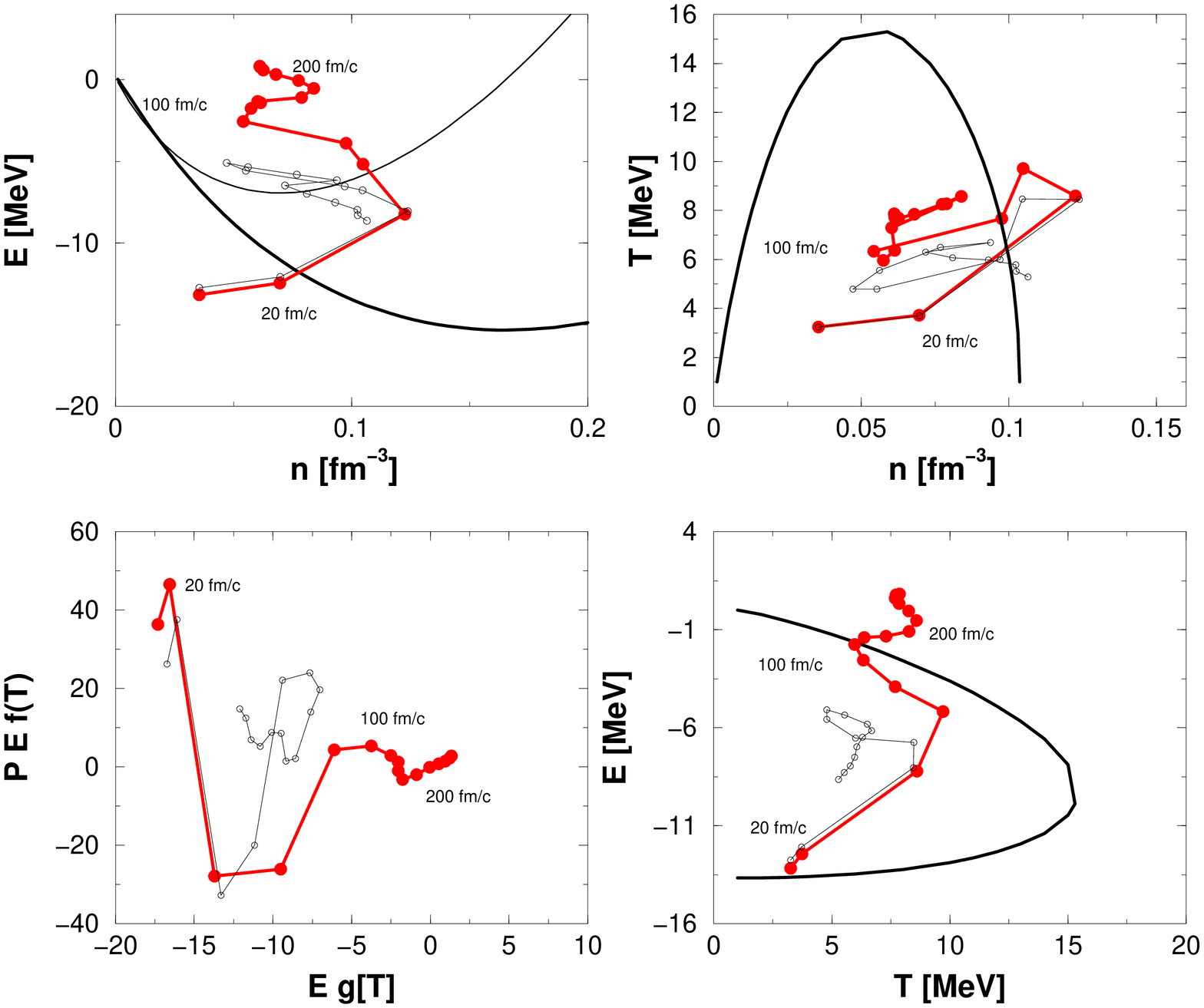,height=10cm,angle=-90}
}
\hspace{0.2cm}
\parbox[]{6.5cm}{
\vspace{7cm}\caption{The same figure as \protect\ref{xesn25} but for $33$MeV lab
  energy. \label{xesn33}}
}}
\end{figure}
If we now plot the same reaction at $50$MeV in figure \ref{xesn50} we
see that the trajectories come at rest outside the spinodal region whatever
plot is used
and no second minima is seen anymore in the iso-nothing plot. But, the
surface emission instability is still very pronounced and is probably here the
leading mechanism of matter disintegration.  
We might now search for a situation where we have the opposite extreme
that is we search for a reaction with as less as possible surface
emission instability and as much as possible spinodal decomposition. For this
reason we might think on asymmetric reactions since the different
sizes of the colliding nuclei might suppress the surface
emission. Indeed as can be seen in figure \ref{niau25} for an
asymmetric reaction of $Ni$ on $Au$ at $25$MeV lab-energy with nearly the same total charge as in the
reaction before that the surface emission instability is less pronounced while
the spinodal instability is much more important. There appears even a
third minima showing that the matter suffers spinodal decomposition
perhaps more than once if the bombarding energy is low enough and a
long oscillating piece of matter is developing.

The higher bombarding energies now show the same qualitative effect
which pronounces the surface emission instability and reduces the importance
of the spinodal decomposition. Much smaller compression densities and
temperatures are reached in these reactions compared to the more symmetric
case of $Xe$ on $Sn$.

\section{Summary}

The extension of BUU simulations by nonlocal shifts and quasiparticle
renormalisation has been presented and compared to recent
experimental data on mid rapidity charge distributions. It is found
that both the nonlocal shifts as well as the quasiparticle renormalisation
must be included in order to get the observed mid--rapidity
matter enhancement.

The inclusion of quasiparticle renormalisation has been performed 
by using the normally  excluded events by Pauli
blocking. Since the quasiparticle renormalisation and corresponding
effective mass features can be considered as zero angle collisions
they can be realized by nonlocal shifts for the scattering events
which are normally rejected. This means that one has to perform the
advection step for the cases of Pauli blocked collisions without
colliding the particles. Besides giving a better description of experiments,
this has the effect of a dynamically softening of equation of state seen
in longer oscillations of giant compressional resonance.

In this way we present a combined picture including nonlocal off-sets
representing the nonlocal character of scattering, which leads to
virial correlations with the quasiparticle renormalisation, and as a
result to mean field fluctuations. We propose that no additional
stochasticity need to be assumed in order to get realistic
fluctuations.

The nonlocal kinetic theory leads to a different nonequilibrium
thermodynamics compared to the local BUU. We see basically a higher
energetic particle spectra and a higher transversal 
temperature of $2$MeV. This is
attributed to the conversion of two-particle correlation energy into
kinetic energy which is of course absent in local BUU scenario.

By constructing a temperature independent combination of
thermodynamical variables we are able to investigate the signals of
phase transitions under iso-nothing conditions. Two mechanisms of instability have been identified:
surface emission instability and spinodal decomposition. We predict for the
currently investigated reactions seen in table \ref{tt}
which effect should be the leading one for matter decomposition.
\begin{table}
\centerline{\parbox[]{14cm}{
\parbox[]{6cm}{
\begin{tabular}[t]{|l ||c|c|c|}
MeV&25&33&50\\
\hline\hline
&&&\\
$^{58}_{28}$Ni$$ + $^{197}_{79}$Au$$&S&CS&C(S)\\&&&\\
$^{129}_{54}$Xe$$ + $^{119}_{50}$Sn$$&CS&C(S)&C\\&&&\\
\end{tabular}
}
\hspace{1cm}
\parbox[]{6cm}{
\begin{tabular}[t]{|l ||c|c|c|}
MeV&15&33&60\\
\hline\hline
&&&\\
$^{157}_{64}$Gd$$ + $^{238}_{92}$U$$  &--  &CS&C\\&&&\\
$^{181}_{73}$Ta$$ + $^{197}_{79}$Au$$ &CS&C(S)&C\\&&&\\
\end{tabular}
}}}
 \caption{The prediction of the leading mechanisms of matter
   disintegration for two reactions with equal total charge but
   symmetric and asymmetric entrance channels. Surface compression is denoted by $C$
   and spinodal decomposition by $S$.\label{tt}}
\end{table}
In the reactions with bombarding energies higher than the Fermi energy the 
fast surface eruption happens outside the spinodal region. 
For even higher energies there is not enough time for the system to rest at
the spinodal region. The trajectories simply move through the spinodal and
the 
system decays before it comes to an equilibrium - like state inside
the spinodal region. Around the Fermi energy the spinodal
decomposition might occur and is accompanied by an anomalous velocity
and density profile \cite{MTP00}.

\section{Acknowledgements}
The authors would like to thank the members of the INDRA collaboration for
providing the experimental data.


\begin{thebibliography}{10}

\bibitem{Stuttge92}
L. Stuttg{\'e} and {\it et~al.}, Nucl. Phys. A {\bf 539},  511  (1992).

\bibitem{SDCD97}
L.~G. Sobotka, J.~F. Dempsey, R.~J. Charity, and P. Danielewicz, Phys. Rev. C
  {\bf 55},  2109  (1997).

\bibitem{PR87}
C. Pethik and D. Ravenhall, Nucl. Phys. A {\bf 471},  19c  (1987).

\bibitem{DDM90}
R. Donangelo, C.~O. Dorso, and H.~D. Marta, Phys. Lett. B {\bf 263},  19
  (1991).

\bibitem{DRS91}
R. Donangelo, A. Romanelli, and A.~C.~S. Schifino, Phys. Lett. B {\bf 263},
  342  (1991).

\bibitem{KH94}
D. Kiderlen and H. Hofmann, Phys. Lett. B {\bf 332},  8  (1994).

\bibitem{ACC95}
S. Ayik, M. Colonna, and P. Chomaz, Phys. Lett. B {\bf 353},  417  (1995).

\bibitem{BV85}
R. Balian and M. Veneroni, Ann. of Phys. {\bf 164},  334  (1985).

\bibitem{Fl89}
H. Flocard, Ann. of Phys. {\bf 191},  382  (1989).

\bibitem{TV85}
T. Troudet and D. Vautherin, Phys. Rev. C {\bf 31},  278  (1985).

\bibitem{ColDiT95}
M. Colonna, M. DiToro, and A. Guarnera, Nucl. Phys. A {\bf 589},  160  (1995).

\bibitem{CTGMZW98}
M. Colonna {\it et~al.}, Nucl. Phys. A {\bf 642},  449  (1998).

\bibitem{FCD98}
G. Fabbri, M. Colonna, and M. DiToro, Phys. Rev. C {\bf 58},  3508  (1998).

\bibitem{BCR91}
G.~F. Burgio, P. Chomaz, and J. Randrup, Nucl. Phys. A {\bf 529},  157  (1991).

\bibitem{SLM96}
V. {\v S}pi{\v c}ka, P. Lipavsk{\'y}, and K. Morawetz, Phys. Lett. A {\bf 240},
   160  (1998).

\bibitem{MLSCN98}
K. Morawetz {\it et~al.}, Phys. Rev. Lett. {\bf 82},  3767  (1999).

\bibitem{LSM99}
P. Lipavsk{\'y}, V. {\v S}pi{\v c}ka, and K. Morawetz, Phys. Rev. E {\bf 59},
  {R 1291}  (1999).

\bibitem{MLSK98}
K. Morawetz, P. Lipavsk{\'y}, V. {\v S}pi{\v c}ka, and N. Kwong, Phys. Rev. C
  {\bf 59},  3052  (1999).

\bibitem{MLNCCT01}
K. Morawetz {\it et~al.}, Phys. Rev. C {\bf 63},  034619  (2001).

\bibitem{B00}
F. Bocage and et~al., Nucl. Phys. A {\bf 676},  391  (2000).

\bibitem{P00}
E. Plagnol and et~al., Phys. Rev. C {\bf 61},  014606  (2000).

\bibitem{G98}
E. Galichet, Ph.D. thesis, Insitut de Physique Nucl\'eaire de Lyon, 1998.

\bibitem{L00}
J.~F. Lecolley and et~al., Nucl. Inst. and Meth. A {\bf 441},  517  (2000).

\bibitem{LSM97}
P. Lipavsk{\'y}, K. Morawetz, and V. {\v S}pi{\v c}ka,   (1999), book in press
  Annales de Physique, K. Morawetz, Habilitation University Rostock 1998.

\bibitem{NTL91}
P.~J. Nacher, G. Tastevin, and F. Lalo{\"e}, Ann. Phys. (Leipzig) {\bf 48},
  149  (1991).

\bibitem{H90}
M. de~Haan, Physica A {\bf 164},  373  (1990).

\bibitem{MT00}
K. Morawetz, Phys. Rev. C {\bf 62},  44606  (2000).

\bibitem{M00}
K. Morawetz, Phys. Rev. C 63, 14609 {\bf 63},  14609  (2001).

\bibitem{MK97}
K. Morawetz and H. K{\"o}hler, Eur. Phys. J. A {\bf 4},  291  (1999).

\bibitem{SHJGRSS89}
P. Schuck {\it et~al.}, Prog. Part. Nucl. Phys. {\bf 22},  181  (1989).

\bibitem{GW00}
T. Gaitanos, H. Wolter, and C. Fuchs,   (2000), sub: nucl-th/0003043.

\bibitem{GWF00}
T. Gaitanos, H. Wolter, and C. Fuchs, Phys. Lett. B {\bf 478},  79  (2000).

\bibitem{HS95}
C.~M. Hung and E.~V. Shuryak, Phys. Rev. Lett. {\bf 75},  4003  (1995).

\bibitem{MTP00}
K. Morawetz, S. Toneev, and M. Ploszajczak, Phys. Rev. C {\bf 62},  64602
  (2000).

\end{thebibliography}

\end{document}